\documentclass[]{spie}  
 
\usepackage{amsmath,amsfonts,amssymb}
\usepackage{graphicx}
\usepackage[labelfont=bf]{caption}

\usepackage{hyperref}
\hypersetup{colorlinks=true,allcolors=[rgb]{0,0,0.6}}

\graphicspath{{figures/}}

\title{Modeling of a stepped Luneburg lens for all-sky surveys}

\author[a]{Mason Carney}
\author[a]{Matthew A. Kenworthy}

\affil[a]{Leiden Observatory, Leiden University, P.O. Box 9513, 2300 RA Leiden, The Netherlands}

\authorinfo{E-mail: \href{mailto:kenworthy@strw.leidenuniv.nl}{kenworthy@strw.leidenuniv.nl}, Telephone: +31 (0) 71 527 8455 \\ The stepped ray trace is publicly available at: \url{https://github.com/masoncarney/stepped_luneburg/}}

\begin{document}
\maketitle

\begin{abstract}

We investigate the scattered light properties of a Luneburg lens approximated as a series of concentric shells with discrete refractive indices.
The stepped Luneburg lens has been previously modeled at microwave wavelengths with full solutions for the electromagnetic field equations when the lens is of comparable size to the wavelength.
We investigate the properties of a Luneburg lens at optical wavelengths using a geometric ray tracing technique.
We develop a stack-based ray tracing algorithm with the \textsc{python} programming language that tracks all reflected and refracted rays generated at each optical interface.
The code shows that a Luneburg lens with 40 steps and a refractive index power-law exponent of 0.55 will produce images of nearly all naked eye ($<$6) magnitude stars with an enclosed energy of 50\% at a spatial resolution of 3.2 degrees.
We find 72 cases of blended stars where a star with magnitude $<$6 falls within 3 degrees of angular separation from a star with magnitude $<1$.
The optical stepped Luneburg lens has promising applications for low-cost, continuous all-sky monitoring to obtain transit light curves of bright, nearby stars.

\end{abstract}


\section{INTRODUCTION}

\label{sec:intro} 

A Luneburg lens is a spherically symmetric optic with a variable index of refraction $n$ that is a function of the radius $r$ such that $n = n(r)$.
The equations that govern the relationship between $n$ and $r$ can give the lens the ability to form perfect geometrical images of two concentric spheres onto one another.
A description of the Luneburg lens was first given by Rudolf Luneburg~\cite{Luneburg44}, where he derived equations for the refractive index $n(r)$ of a spherical lens with two external foci, which gives a maximum refractive index at the lens center that decreases radially outward to the lens surface.
If one of the concentric spheres is at infinity, light rays originating from the outer sphere will produce parallel rays incident on the inner sphere that are then focused to a point on the opposite surface of the inner sphere, given the correct description for the gradient of the refractive index $n(r)$. 
Some applications for a lens with these properties include communication antennas\cite{Walter60,Scheel70, Higgs84} and optical waveguides\cite{Zernike74, Southwell77, Sochacki82}.
The behavior of a Luneburg lens also has benefits for small scale astronomical observations.
Under good observing conditions the Luneburg lens can create an image of the night sky over 2$\pi$ steradians with no active mechanical repointing required.
We aim to test the feasibility of the Luneburg lens for small scale observations, such as those done at a high school or university, or for outreach and educational purposes.
As a passive instrument with a wide field of view, it would be a useful tool for low-budget astronomy operations if the lens can be manufactured at a reasonable cost.
In this paper we consider limitations regarding the production and performance of a Luneburg lens and optimize the lens characteristics for the best image quality.
A continuously varying index of refraction is not practical to construct, but the Luneburg lens can be approximated with a series of discrete layers, each with its own refractive index.
We explore the number of discrete lens layers needed for adequate imaging of naked eye stars and determine the quality of an image at the focal surface using the enclosed intensity as our metric.
%

\section{DESCRIPTION OF STEPPED LUNEBURG LENS} 

To construct a model for all-sky imaging, we are concerned only with the two external foci case for the Luneburg lens\cite{Morgan58} with an index of refraction $n_{\rm L}(r)$ described by: 

\begin{align}
 &n_{\rm L} = \exp[\omega(\rho,r_0) + \omega(\rho,r_1)] \notag \ {\rm and}\\
 &\rho = r n_{\rm L} \ {\rm for} \ 0 \geq \rho \geq 1,
 \label{Lun_gen}
\end{align}

\noindent where $r_0$ and $r_1$ indicate the location of the focal point of each sphere. We set $r_1 = \infty$ for imaging stars in the sky.
Equation~\ref{Lun_gen} was derived under the condition that the radius of the inner sphere is set to unity.
Since incoming rays originate at infinity, they should focus to a point on the surface of the lens at $r=1$, thus we set $r_0 = 1$.
The function $\omega$ at $r_0 = 1$ and $r_1 = \infty$ is described by:

\begin{align}
 &\omega(\rho,1) = \frac{1}{2} \log[1 + (1-\rho^{2})^{1/2}], \notag \\
 &\omega(\rho,\infty) = 0.
 \label{omega_spec}
\end{align}

Combining Equation~\ref{Lun_gen} with Equation~\ref{omega_spec}, we obtain the simplified function $n(r)$ for the refractive index of a Luneburg lens with radius set to unity:

\begin{equation}
 n = (2-r^{2})^{1/2}.
 \label{Lun_simp}
\end{equation}

Since we are modeling a Luneburg lens in discrete steps rather than a continuously variable refractive index, Equation~\ref{Lun_simp} gives us the refractive index for a shell of a given radius within the lens.
At the surface of each lens layer a ray of light will encounter a refractive index transition and its behavior will be described by the vector form of Snell's law, which is based on the light ray direction vector $\vec{d}$ and the normal vector of the lens layer $\vec{n}$, both normalized to unity.

\begin{align}
 &\cos{\theta_1} = \vec{n} \cdot (-\vec{d}) \notag \\
 &\cos{\theta_2} = \sqrt{1 - \left(\frac{n_1}{n_2}\right)^{2} (1-(\cos{\theta_1})^{2})} \notag \\
 &\vec{\mathrm{v}}_{\mathrm{reflected}} = \vec{d} + (2 \cos{\theta_1}) \vec{n} \notag \\
 &\vec{\mathrm{v}}_{\mathrm{refracted}} = \left(\frac{n_1}{n_2}\right) \vec{d} + \left(\frac{n_1}{n_2} \cos{\theta_1} - \cos{\theta_2}\right) \notag \\
 &\mathrm{where} \, \, \cos{\theta_1} \geq 0. 
 \label{Snell}
\end{align}

With Equation~\ref{Snell} we are able to calculate the reflected and refracted rays produced for each ray-layer intersection until all rays have exited the lens.
The result is an image at the focal point surrounded by an accumulation of extraneous rays due to the discrete nature of the lens.
Analysis of the image produced at the lens focal point and halos allows us to determine the resolution and overall image quality for a stepped Luneburg lens.
%

\section{STACK RAY TRACE CODE}

In order to determine the scattered light properties of a stepped Luneburg lens, we have written a stack-based ray tracing code that can propagate the large number of reflected and refracted rays generated from one incident ray.
With this code we are able to accurately propagate thousands of rays and track the position, direction, and amplitude of each ray individually.
The code is written for three-dimensional ray propagation so that manufacturing errors, such as the decentering of layers or deviations from the ideal spherical layer may be introduced.

\begin{figure}[!b]
\centering
\includegraphics[width=10cm,height=10cm,keepaspectratio=true]{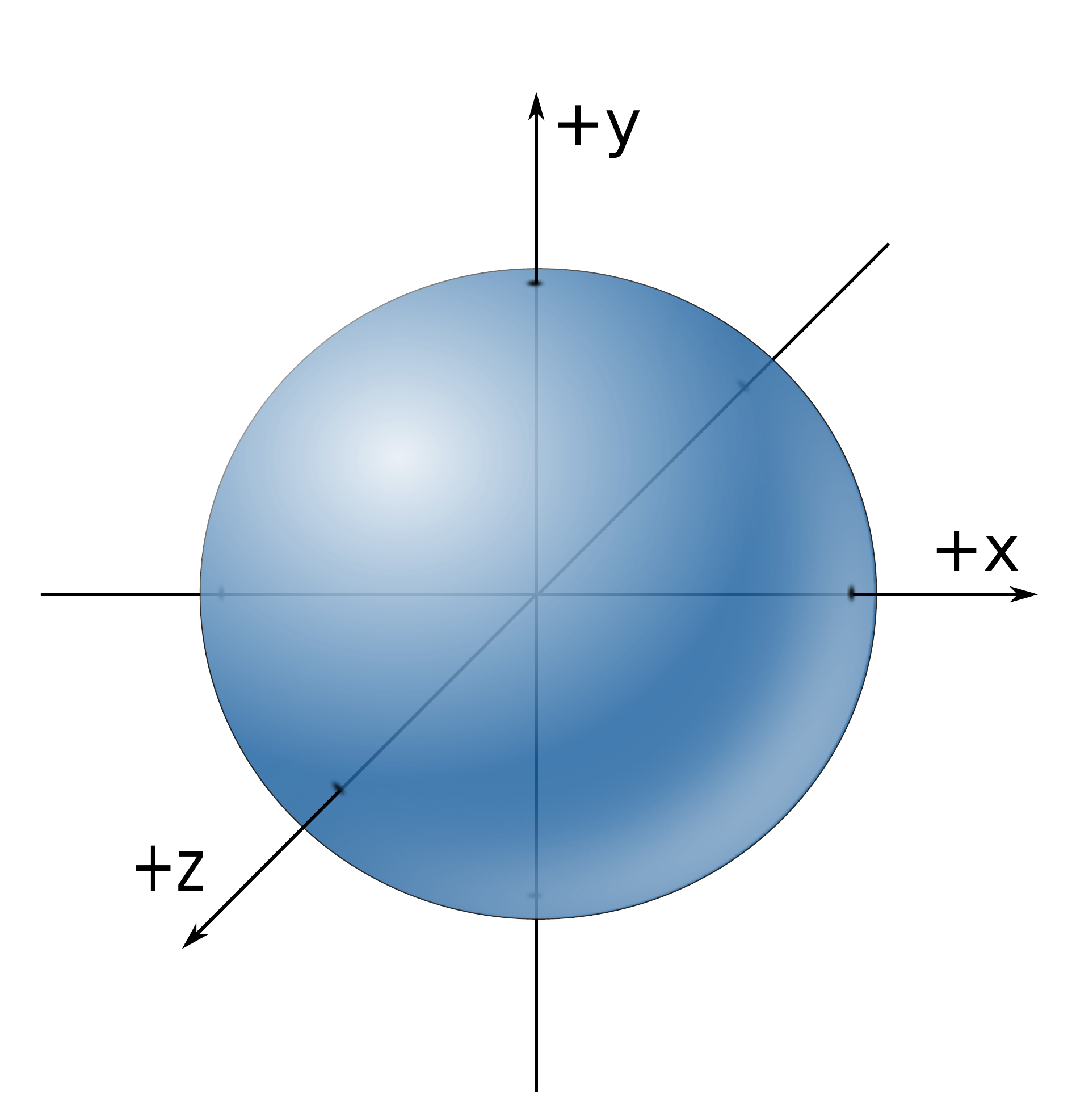}
\caption{\small{Orientation of the Luneburg lens model in three-dimensional $(x,y,z)$ coordinates.} }
\label{fig_orientation}
\end{figure}

Construction of a Luneburg lens first requires the creation of a multi-layer sphere in three-dimensional space.
The origin of our coordinate system is designated as $(0,0,0)$ in a Cartesian $(x,y,z)$ system and the axes are chosen such that the $y$-axis runs through the zenith and nadir of the lens, while the top and bottom hemispheres of the lens are divided by the $x-z$ plane (see Figure~\ref{fig_orientation}). 
The number of discrete lens layers can be adjusted, but the outermost radius is normalized to unity for the purpose of calculating the Luneburg refractive indices in Equation~\ref{Lun_simp}.
%


The code initializes input arrays for a wavefront of light rays located above the lens and stores the position, direction, and amplitude of all incoming light rays. Each initial input ray is given an amplitude equal to 1. 
The input arrays can be constructed in four ways: a line of rays along the $x$-axis at $y=2$, a uniform square grid of rays in the $x-z$ plane from $(-1,-1)$ to $(1,1)$ at $y=2$, a random grid of rays in the $x-z$ plane from $(-1,-1)$ to $(1,1)$ at $y=2$, and a user-provided star pattern where a uniform square grid of rays at $y=2$ is rotated to the three-dimensional position of each star such the focal points of the wavefronts recover the star pattern on the lens bottom surface.
The tracing of each ray is done as part of a stacked, iterative loop process, where ray properties are extracted for one ray at a time and the direction of the ray determines whether or not that ray intersects with a layer of the lens.
We define the ray properties in vector form, thus we can test for the intersection between a ray and a lens interface using a ray described by the equation:

\begin{equation}
 \vec{p} = t\vec{d} + \vec{p}_0,
 \label{line}
\end{equation}

\noindent where $\vec{p}$ is a point $(x,y,z)$, $t$ is a parametric variable, $\vec{d}$ is the direction vector, and $\vec{p}_0$ is the original $(x,y,z)$ point of the ray. We can insert the ray vector into the equation for a sphere and get the following quadratic equation:

\begin{align}
 &x^{2} + y^{2} + z^{2} = \vec{p} \cdot \vec{p} - r^{2} = 0, \notag \\
 &(t\vec{d} + \vec{p}_0 - \vec{p}_c) \cdot (t\vec{d} + \vec{p}_0 - \vec{p}_c) - r^{2} = 0,
 \label{eq_sphere}
\end{align}

where $\vec{p}_c$ is the center of the lens.
The quadratic formula can be used to solve for $t$, and the sign of the discriminant can be used to test for the intersection between the ray and the lens interface:

\begin{align}
 &b^{2} - 4 a c \notag \\
 & \notag \\
 &a = \vec{d} \cdot \vec{d} \notag \\
 &b = 2 \vec{d} \cdot (\vec{p}_0 - \vec{p}_c) \notag \\
 &c = (\vec{p}_0 - \vec{p}_c) \cdot (\vec{p}_0 - \vec{p}_c) - r^{2},
 \label{discr}
\end{align}

If both solutions to the discriminant in Equation~\ref{discr} are negative, no intersection occurs.
If a there is at least one positive value, the ray will intersect the lens interface.
We take the smallest value of the solutions for $t$ because this corresponds to an intersection between the ray and the closest lens interface. 

\begin{table}[!ht]
\begin{tabular}{l}
\textbf{Stack-based Ray Trace Routine} \\ \hline
\\
1. Input array of position, direction, amplitude for incoming light rays \\ 
2. Calculate intersections of rays and lens layer interface \\ 
3. Implement Snell's law for all ray intersections \\ 
4. Store position, direction, amplitude of reflected and refracted rays \\ 
5. Discard rays that fall below the amplitude cutoff or exit the top of the lens \\
6. Use reflected and refracted rays as input for next iteration of stack \\
7. Go to Step 1 until all rays have exited the lens \\
\end{tabular}
\caption{The structure of the Ray Trace Algorithm}
\label{tab_stacksteps}
\end{table}

In Table~\ref{tab_stacksteps} the algorithm loop is detailed.
The first loop of the stacked, iterative ray tracing routine is a test for intersection with the outermost surface of the lens (i.e. a test to see if the ray will enter the lens).
All incoming rays that do not intersect the lens surface are dropped from the routine.
If a ray enters into the bottom hemisphere of the lens or if it exits through the top hemisphere of the lens, it is also dropped.
If the lens is to be manufactured, only the top hemisphere would be available for incident light while the bottom hemisphere is for detection and imaging.
For rays that do enter the lens, the resulting direction vector is calculated using Equation~\ref{Snell} at the lens surface.
Each implementation of Snell's law calculates a new value for the amplitude of the reflected and refracted rays.
If at any point the amplitude of a ray falls below a preset amplitude threshold, it is dropped from the routine.
We set the amplitude threshold at 0.01.
This avoids the accumulation of large numbers of rays within the lens that contribute little to the overall output intensity.
With a sufficient number of iterations, each ray will either be dropped due to low amplitude or exit the lens.
For all rays that exit the bottom hemisphere of the lens, the position, direction, and amplitude for is recorded analysis.
To test for exiting rays, we consider the dot product of a ray $(x,y,z)$ point and a ray direction vector that intersects the outermost layer of the lens.

\begin{equation}
 \vec{p} \cdot \vec{d} \, \, \, \mathrm{for} \, \, \, r = 1.
 \label{dot_pr}
\end{equation}

If Equation~\ref{dot_pr} is negative for the first loop iteration, then the ray enters the lens.
For positive dot products at the outermost layer, the ray exits the lens.
The position, direction, and amplitude of bottom-exiting rays are stored in preparation for two-dimensional mapping and graphical analysis of the resulting images that form on the bottom hemisphere of the Luneburg lens.

\section{RESULTS FROM SIMULATIONS}

\subsection{Mapping Luneburg lens ray tracing}


%
\begin{figure}[!b]
\centering
\includegraphics[width=13cm,height=13cm,keepaspectratio=true]{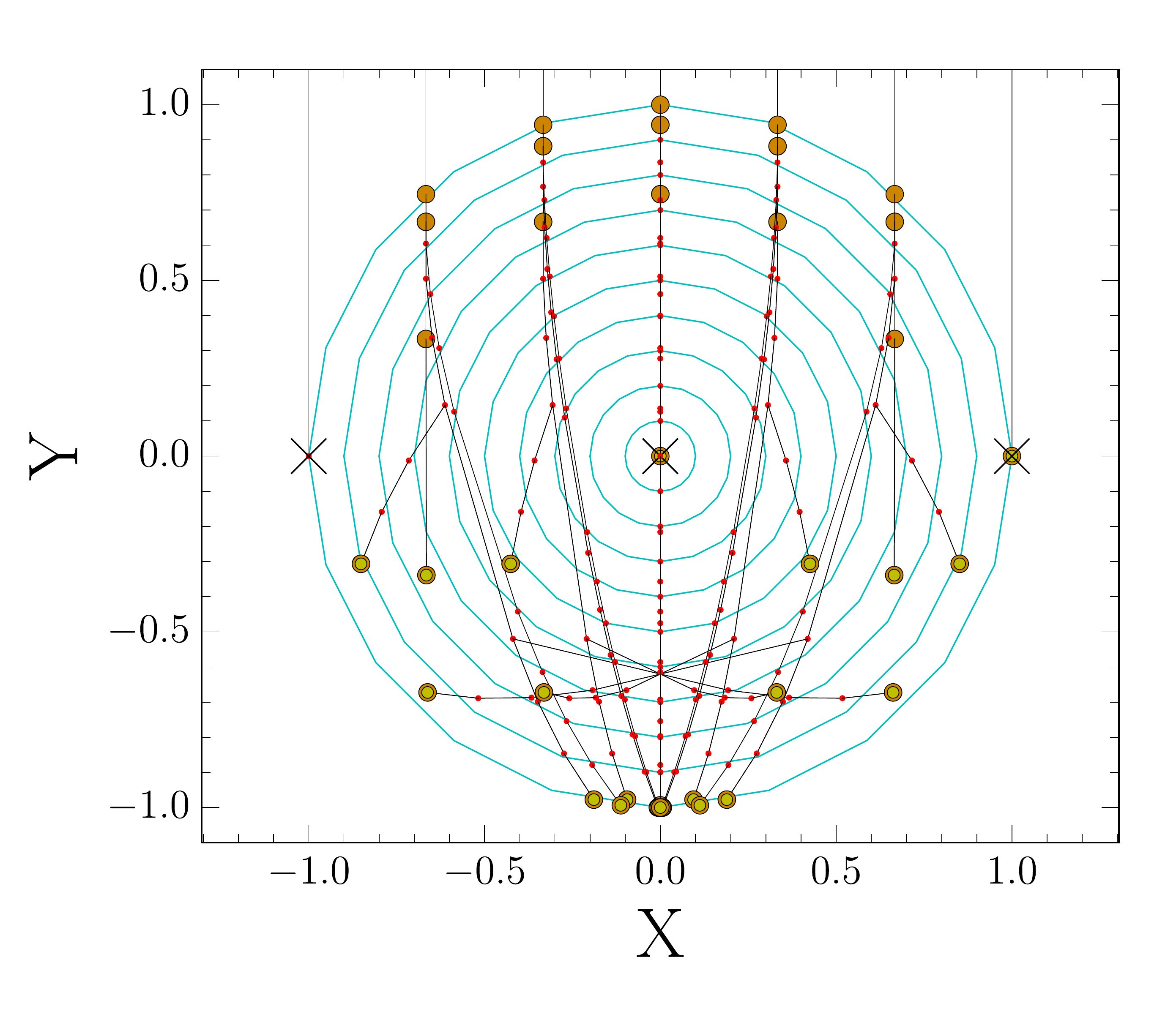}
\caption{\small{Tracking rays through the Luneburg lens with the stack-based ray tracing routine. The plot shows the trajectory of incident rays and trajectories of reflected and refracted rays (black), intersections between a ray and a lens interface (red dots), rays that intersect the lens surface (orange dots), and rays that exit the lens (yellow dots). Rays marked with a black X failed to enter/exit the lens properly.}}
\label{fig_raytracebig}
\end{figure}

A toy model run to showcase the ray tracing code is shown in Figure~\ref{fig_raytracebig} for a lens with 10 layers and 49 rays initialized in a uniform square grid pattern.
The routine tracks the progress of each ray in three-dimensions as it travels through the lens and plots the rays from a side view of the lens in the $x-y$ plane.
Each black line indicates the trajectory of a ray between consecutive intersections with a surface of the lens.
Red dots indicate a ray intersecting a lens interface.
When a ray intersects the outermost surface of the lens, the intersection point is marked with an orange circle.
When a ray exits the lens the intersection point is additionally marked with a smaller yellow circle.
Any point marked with an X indicates that the ray did not properly enter or exit the lens.
Either the ray was incident on the bottom hemisphere, or it exited from the top hemisphere, and therefore must be discarded before further analysis.
This graphic representation of ray trajectories allows us to track each ray individually by visual inspection to confirm that surface interactions are properly implementing Snell's law.
Additionally, we are able to see any potential extraneous rays that might be created due to large angles of incidence at a lens interface. 

To plot the image produced by rays exiting the bottom hemisphere of the lens, the three-dimensional position of each exiting ray is projected onto a two-dimensional plot and weighted with the square of the amplitude to display the intensity associated with each ray.
In this way we are able to create a two-dimensional map of the bottom hemisphere of the lens and obtain a qualitative representation of the position and intensity of the resulting image.
We are able to infer the position of halos around the focal point that are due the discrete lens layers. 
For a more quantitative examination of the image made on the bottom hemisphere of the lens, we generate a pixelated intensity map.
The intensity map is a two-dimensional projection of the three-dimensional positions and the intensities that are stored when the rays exit the bottom of the lens.
The map is centered on the focal point of incident wavefront of light rays and plots a central region that contains a user-provided fraction of the total output intensity. The central region is then normalized to 1 in order to show the relative intensity of the surrounding halos.
Halos with sufficient intensity will cause contamination in the case of multiple bright sources and will distort the final image, reducing the quality of the Luneburg lens.

\subsection{Optimizing lens parameters}

\begin{figure}[!t]
\centering
\includegraphics[width=13cm,height=13cm,keepaspectratio=true]{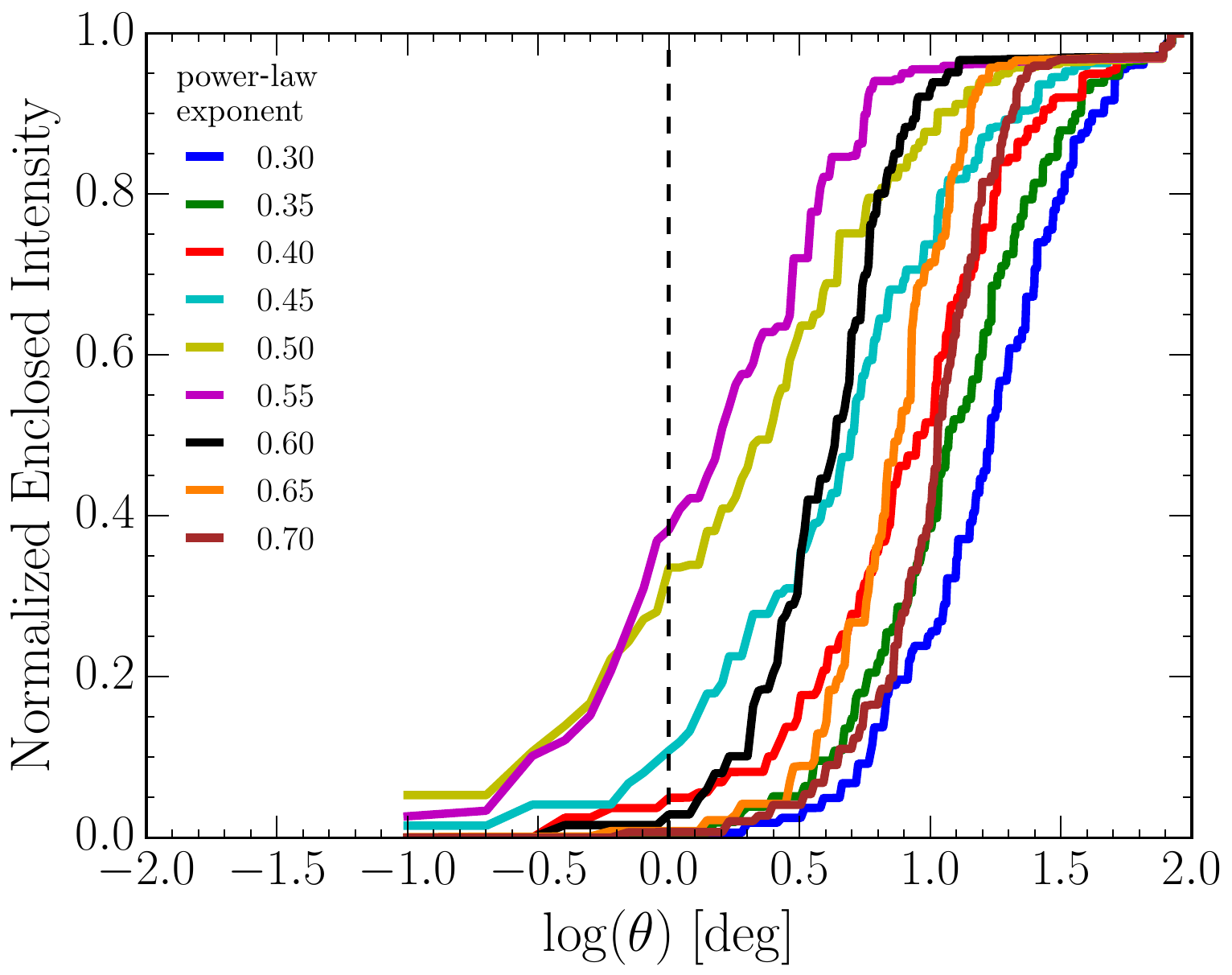}
\caption{\small{The enclosed (cumulative) intensity of exit rays versus the angle $\theta$ away from the focal point of the initial wavefront for a 40 layer Luneburg lens with 1024 rays. $\theta$ is plotted in log scale to emphasize the region close to the focal point. Colored lines represent the power-law exponent $i$ in Equation~\ref{Lun_adj}.}}
\label{fig_encint}
\end{figure}

For analysis of the Luneburg lens imaging quality we use a wavefront of 1024 incident parallel rays uniformly gridded in the $x-z$ plane as the initial conditions of a representative model for observing a single star.
To obtain the best image quality for the lens we consider the effects of changing the number of lens layers and adjusting the exponent of the Luneburg power law in Equation~\ref{Lun_simp} such that we can manually designate a different exponent $i$ to create a new equation for refractive indices.

\begin{equation}
 n = (2-r^{2})^{i}.
 \label{Lun_adj}
\end{equation}

Our chosen metric for image quality is $\theta_{50\%}$, the angular radius of a circle centered on the chief ray within which 50\% of the total output intensity is contained.
We adjust the other parameters of the lens to minimize this angle.
The enclosed intensity as a function of the angle away from the focal point is analyzed for multiple values of the power-law exponent $i$ over the range 0.30 $\leq i \leq$ 0.70.
In Figure~\ref{fig_encint} it is clear that $i = 0.55$ provides the best image quality in the 40 layer case, as it contains the most intensity at small angles away from the focus.
At larger angles we expect the enclosed intensity for all values of $i$ to follow roughly the same trend since the intensity contribution from the outer halos is due mostly to extraneous rays that are not sufficiently refracted through the lens and exit quickly without intersecting many lens interfaces.
We test the same range of power-law exponents for 5 layer, 10 layer, 20 layer, and 40 layer configurations of the lens.
For each layer configuration, we calculate $\theta_{50\%}$ and record the value for each of the Luneburg power-law exponents.
In this way we are able to use both parameters in a single analysis to determine which combination of lens layers and power-law exponents will result in the smallest angle for 50\% enclosed intensity. 

\begin{figure}[!b]
\centering
\includegraphics[width=13cm,height=13cm,keepaspectratio=true]{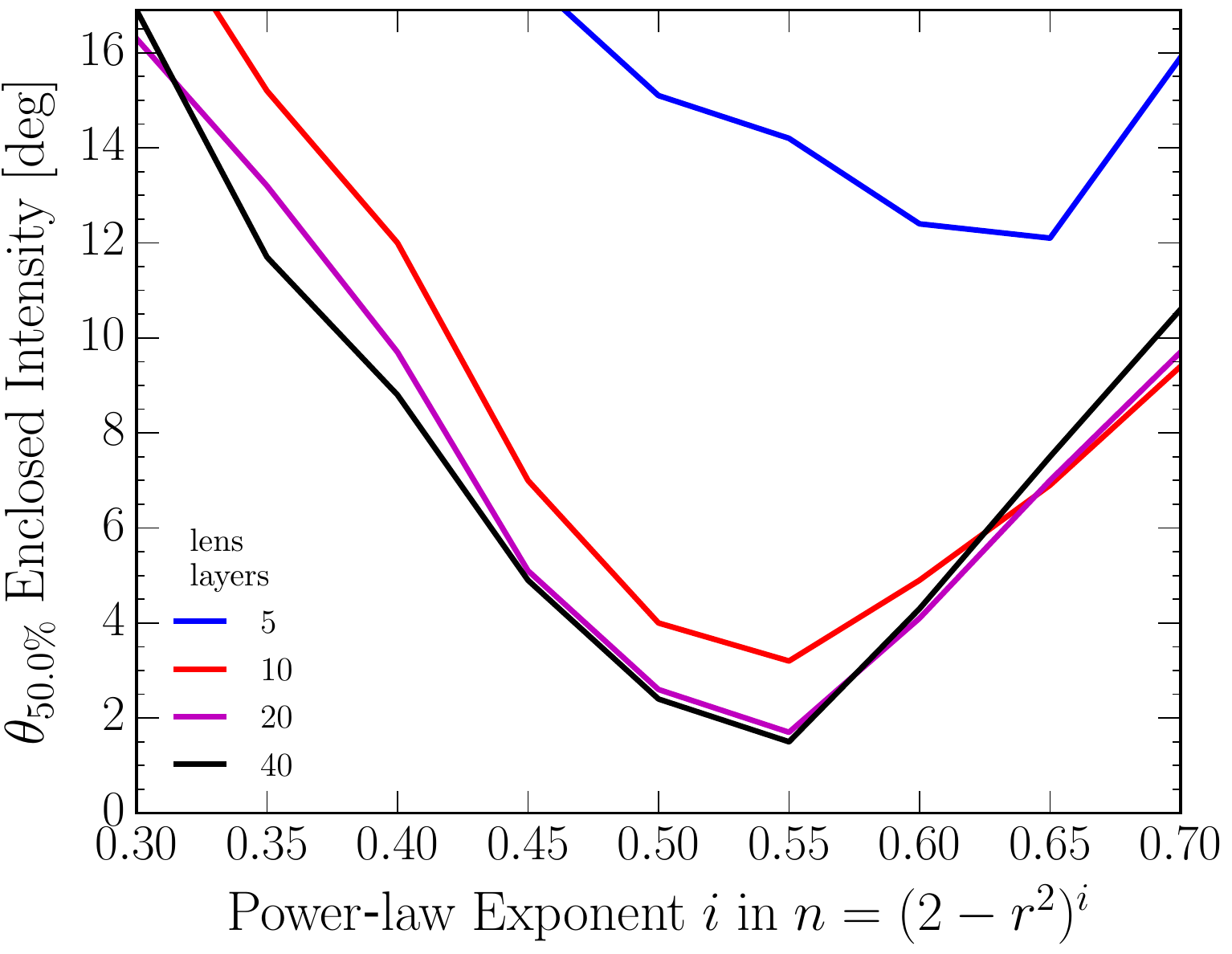}
\caption{\small{The angle $\theta_{50\%}$ containing 50\% of the total output intensity of exit rays versus the power-law exponent $i$ (Equation~\ref{Lun_adj}) for models with 1024 rays. The colored lines represent different numbers of lens layers.} }
\label{fig_angexp}
\end{figure}

From Figure~\ref{fig_angexp} we can see that for increasing lens layers the optimum angle improves as the Luneburg power law tends toward $i = 0.55$. 
While the 40 layer configuration provides the sharpest focusing power, the 20 layer configuration performs nearly as well for $i=0.55$.
These results are consistent with the expectation that as the number of discrete layers increases and the refractive index $n(r)$ for the lens approaches a continuous gradient, the focusing power should improve. We find that the mathematical description for $n(r)$ is best for $i=0.55$ in Equation~\ref{Lun_adj}, which differs from the theoretical prediction by Luneburg in Equation~\ref{Lun_simp}.
We also test the effect of the amplitude cutoff against the completeness of the ray tracing routine.
Snell's law is implemented when a ray intersects and lens interface. After the implementation of Snell's law, if the amplitude of the reflected ray or the refracted ray falls below 1\% of the original ray prior to Snell's law, then the ray is dropped from the ray tracing routine.
An amplitude limit of 1\% corresponds to an intensity limit of 0.01\%.
For the 40 layer Luneburg lens configuration, we set the amplitude cutoff at 10\%, 1\%, and 0.1\% and run the model for multiple values of the power-law exponent $i$ over the range 0.30 $\leq i \leq$ 0.70.

\begin{figure}[!t]
\centering
\includegraphics[width=13cm,height=13cm,keepaspectratio=true]{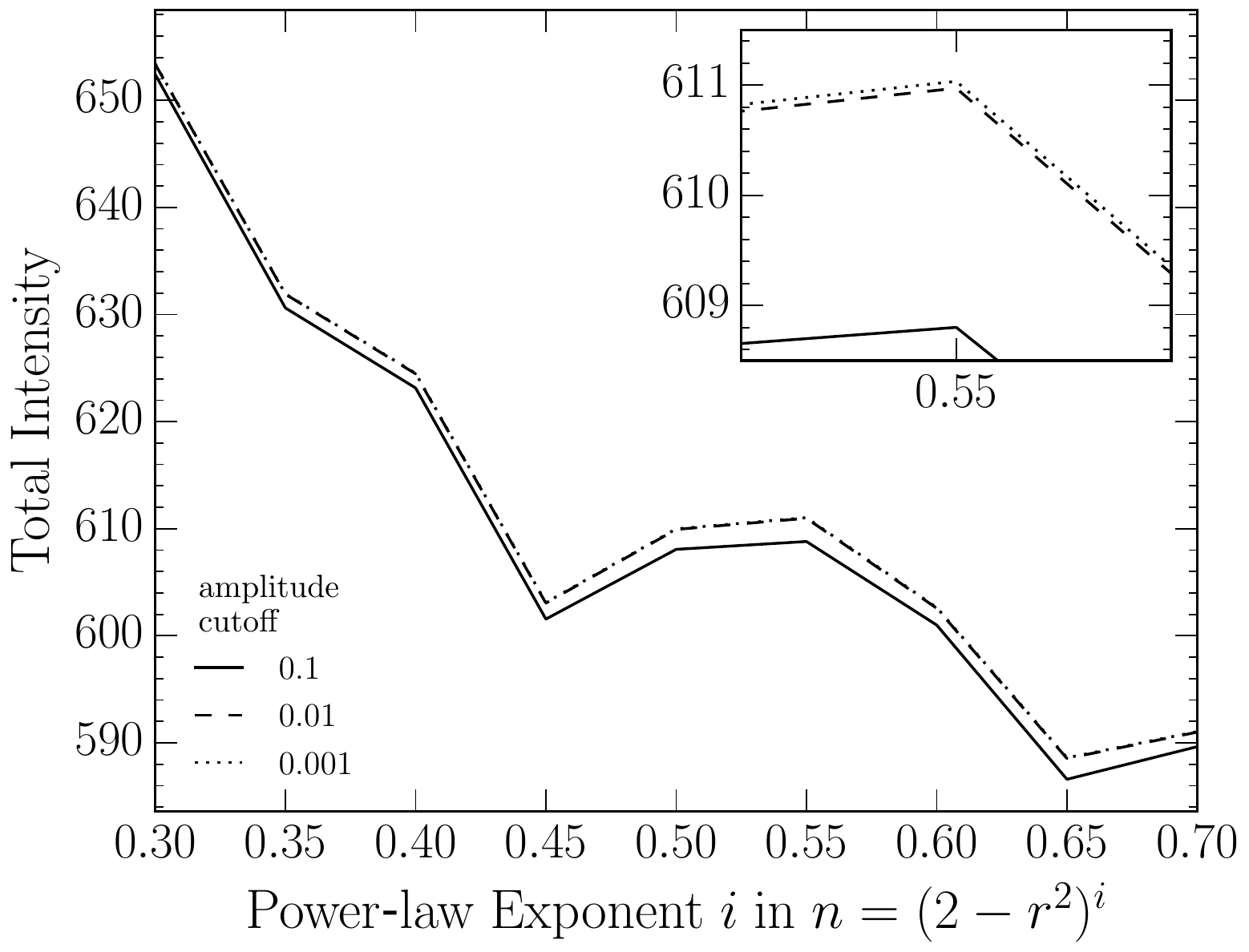}
\caption{\small{The total output intensity of bottom-exiting rays versus the power-law exponent $i$ (Equation~\ref{Lun_adj}) for a 40 layer Luneburg lens model with 1024 rays, where the three lines represent amplitude cutoffs of 10\% (solid), 1\% (dashed), and 0.1\% (dotted). The inset shows a zoomed region for $i=0.55$. There is little improvement between 1\% and 0.1\% amplitude cutoffs. Note that the $y$-axis values differ from the raw amplitude values listed in Table~\ref{tab_eff} because here we plot ray intensity rather than amplitude.} }
\label{fig_angexp_1pct}
\end{figure}

In Figure~\ref{fig_angexp_1pct} we plot the total intensity output of the Luneburg lens versus the power-law exponent $i$ for the three different amplitude cutoff values. There is a significant improvement between the 10\% and 1\% amplitude cutoffs for most values of $i$. Lowering the amplitude cutoff to 0.1\% results in little improvement but a significant increase in the runtime of the code. Table~\ref{tab_eff} compares the amplitude completion of the ray tracing routine for the 40 layer lens configuration.
\begin{table}[!b]
\begin{center}
\resizebox{14cm}{!}{
\begin{tabular}{lrrrrr}
\multicolumn{6}{c}{Ray amplitude completion in the Luneburg lens model} \\ \hline \hline
Cutoff & Total input & Total output & Completion & \multicolumn{2}{c}{Amplitude dropped} \\ \hline
& & & & Due to cutoff & Exit top hemisphere \\ 
10.0\% & 740.00 & 669.44 & 90.5\% & 70.56 & 0.00 \\ 
1.0\% & 740.00 & 710.34 & 96.0\% & 26.28 & 3.38 \\ 
0.1\% & 740.00 & 724.31 & 97.9\% & 9.92 & 5.77 \\ \hline
\end{tabular}
}
\end{center}
\caption{Amplitude completion of a 40 layer Luneburg lens with respect to the total amplitude of rays incident on the lens.
Calculations are made with a Luneburg power-law exponent $i$ = 0.55 and 1024 rays in the initial wavefront. Note that not all rays enter the lens in the uniformly gridded initial wavefront.}
\label{tab_eff}
\end{table}
The effect of decreasing the amplitude cutoff is significant for a reduction from 10\% to 1\%, where we see the completion increase from 90.5\% to 96.0\%. The completion only improves by 2\% when the amplitude cutoff is reduced from 1\% to 0.1\%.
We conclude that there is little improvement for amplitude cutoff values below 1\%, and we can successfully model a Luneburg lens with a 1\% amplitude cutoff without loss of significant ray data.
The analysis performed in this section suggests that the 40 layer Luneburg lens with a power-law exponent $i = 0.55$ from Equation~\ref{Lun_adj} is the optimal set of parameters for best image quality.

\subsection{Observing Quality}

We are able to determine the resolving power of the Luneburg lens using $\theta_{50\%}$.
Analysis of the 40 layer lens configuration with power-law exponent $i=0.55$ and 1024 uniformly gridded initial input rays gives a value $\theta_{50\%}$ = 1.6 degrees.
This is the angle as measured from the center of the focal point, meaning that the value must be doubled to obtain the lens imaging resolution, giving the optimized Luneburg lens an angular resolution of $\theta_{res}$ = 3.2 degrees.
The lens should be capable of resolving all bright stars that are separated by more than 3.2 degrees on the sky.
The central region and any rings containing some fraction of the output intensity must be corrected for the area of the region. 
We calculate the area of the central region as $\pi r_{50\%}^{2}$ and the area of each ring as $\pi(r_{outer}^{2} - r_{inner}^{2})$ and proceed to divide the summed intensity in each region by the area in which they are contained.
After correcting for area, we further normalize by dividing each ring by the area-corrected intensity of the central region out to $r_{50\%}$.
By doing so, we are able to use the central region containing 50\% of the total output intensity as a normalized reference point and can investigate the halo intensities relative to this central point.

\begin{figure}[!b]
\centering
\includegraphics[width=8cm,height=8cm,keepaspectratio=true]{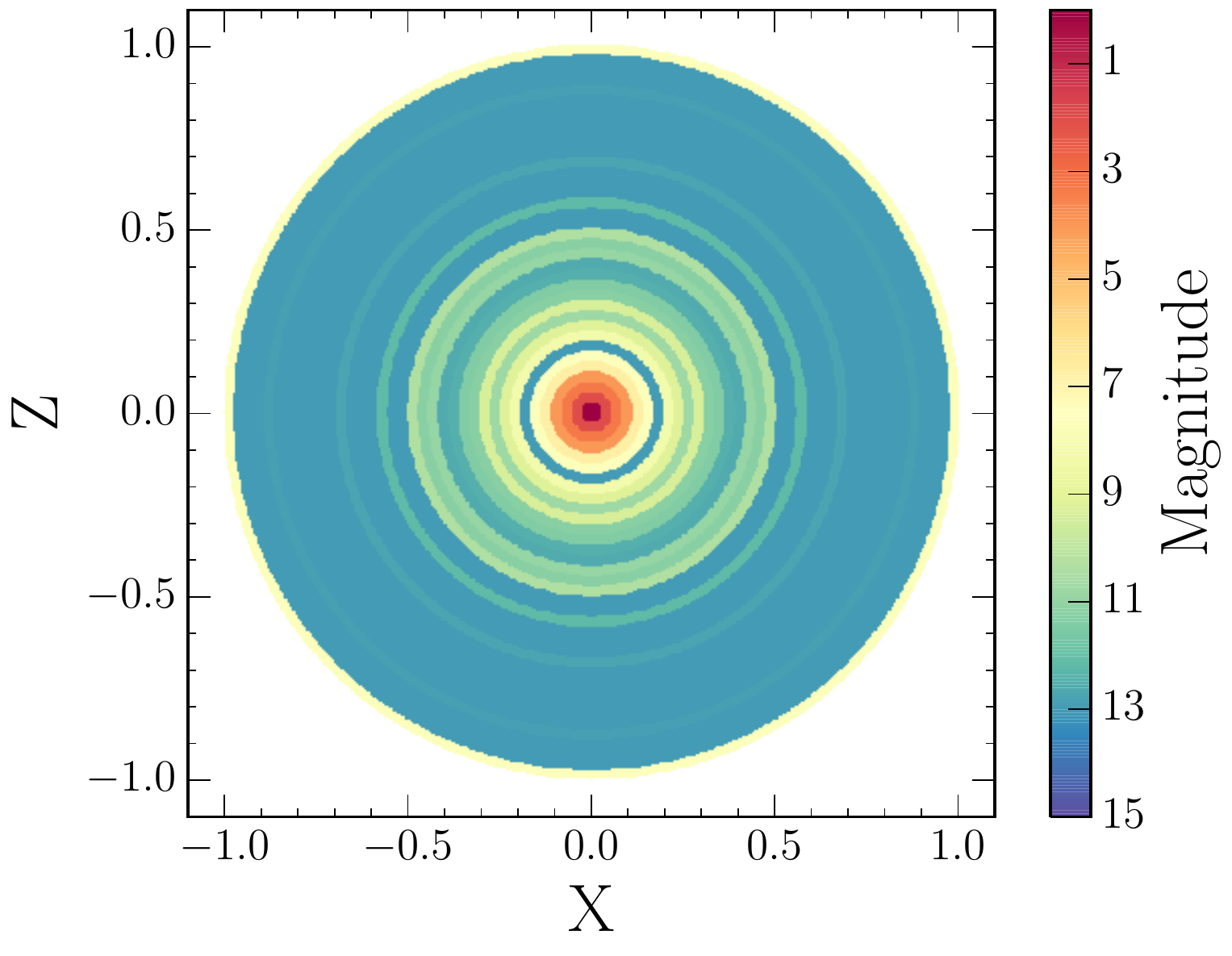}
\includegraphics[width=8cm,height=8cm,keepaspectratio=true]{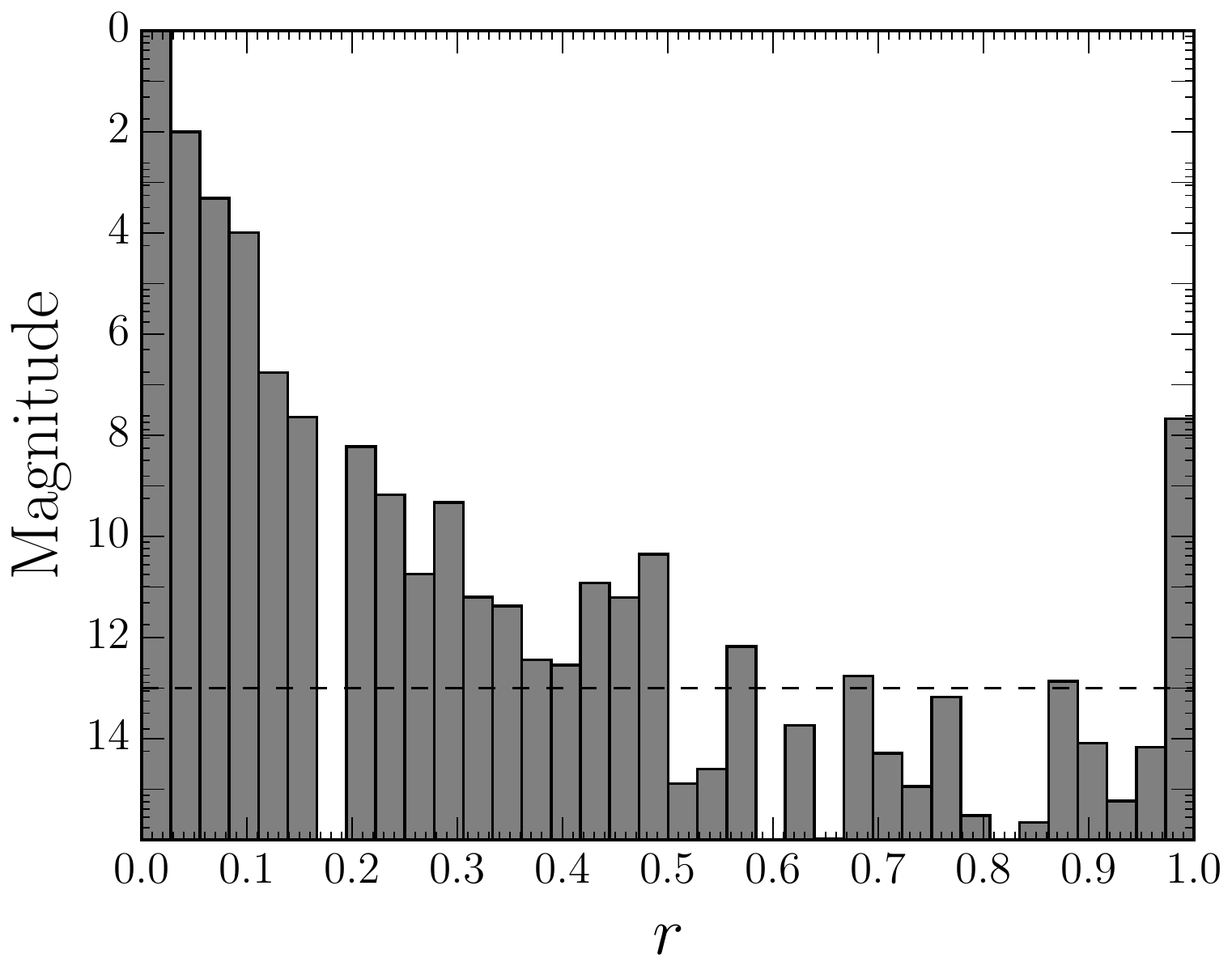}
\caption{\small{({\it Left}) Pixelated intensity map showing the relative intensity of halos due to the discrete nature of the stepped Luneburg lens. halos are plotted in magnitudes with respect to a central region enclosing 50\% of the output intensity for a 40 layer Luneburg lens with power-law exponent $i=0.55$ and 1024 rays. ({\it Right}) Histogram showing the location and relative magnitude of halos.
The dashed black line indicates the sky brightness background.} }
\label{fig_imgmaphist}
\end{figure}

In Figure~\ref{fig_imgmaphist} we see that the majority of halos have very low relative intensity.
Beyond the normalized $r_{50\%}$ central region, most halos range from $4--15$ magnitudes, which are unlikely to diminish imaging quality by overlap or smearing.
Halos from the first $2--3$ rings are stronger, with magnitudes $2--4$ times greater than the central region, which may affect the image quality if there are other nearby bright stars.
Any halos with magnitude $>$13 are below our designated sky brightness background and would not affect observations.

\subsection{Feasibility analysis}

When considering the observational limitations of the Luneburg lens, we are interested in star proximity and the ability of the lens to properly resolve stars that are sufficiently close to one another.
With an image resolution of $\theta_{res}$ = 3.2 degrees we expect to lose resolving capability for stars that are separated by less than 3.2 degrees on the sky.
Our observations focus on the night sky as seen by the naked eye, therefore we do not consider stars that have apparent magnitudes $>6$.
Of particular interest is the separation of $<$6 magnitude stars with respect to the very brightest stars in the night sky.
A bright star in close proximity to other stars will be the most difficult to resolve, as the high degree of stellar brightness will create strong halos that diminish resolving ability in addition to the close angular separation.
Using the Yale Bright Star Catalog\cite{Hoffleit91}, we compile a list of all stars with magnitude $<$1 and search for all stars with magnitude $<$6 that are within 3 degrees of angular separation from the magnitude $<$1 stars.

\begin{table}[!t]
\begin{center}
\begin{tabular}{llrrrc}
\multicolumn{6}{c}{Close-proximity bright stars} \\ \hline \hline
Name & HD identifier & RA (J2000) & DEC (J2000) & $V_{\rm mag}$ & Stellar neighbors \\ 
 &  & [hh;mm;ss]  & [dd;mm;ss]  &  &  $V_{\rm mag} < 6$; $\theta_{\rm sep} < 3^{\circ}$ \\ \hline
Alp Eri & 10144 & 01 37 42.9 & -57 14 12 & 0.46 & 2 \\ 
87 Alp Tau & 29139 & 04 35 55.2 & +16 30 33 & 0.85 & 15 \\ 
13 Alp Aur & 34029 & 05 16 41.4 & +45 59 53 & 0.08 & 1 \\ 
19 Bet Ori & 34085 & 05 14 32.3 & -08 12 06 & 0.12 & 6 \\ 
58 Alp Ori & 39801 & 05 55 10.3 & +07 24 25 & 0.50 & 5 \\ 
Alp Car & 45348 & 06 23 57.1 & -52 41 45 & -0.72 & 4 \\ 
9 Alp CMa & 48915 & 06 45 08.9 & -16 42 58 & -1.46 & 8 \\ 
10 Alp CMi & 61421 & 07 39 18.1 & +05 13 30 & 0.38 & 4 \\ 
67 Alp Vir & 116658 & 13 25 11.6 & -11 09 41 & 0.98 & 2 \\ 
Bet Cen & 122451 & 14 03 49.4 & -60 22 23 & 0.61 & 1 \\ 
16 Alp Boo & 124897 & 14 15 39.7 & +19 10 57 & -0.04 & 2 \\ 
Alp 1 Cen & 128620 & 14 39 35.9 & -60 50 07 & -0.01 & 5 \\ 
21 Alp Sco & 148478 & 16 29 24.4 & -26 25 55 & 0.96 & 3 \\ 
3 Alp Lyr & 172167 & 18 36 56.3 & +38 47 01 & 0.03 & 6 \\ 
53 Alp Aql & 187642 & 19 50 47.0 & +08 52 06 & 0.77 & 8 \\ \hline
\end{tabular}
\end{center}
\caption{List of all stars with magnitude $<$1 taken from the Yale Bright Star Catalog\cite{Hoffleit91} and their $<$6 magnitude neighbors within 3 degrees angular separation.}
\label{tab_brightstars}
\end{table}

For the 15 brightest stars with magnitude $<$1, there are a total of 72 neighboring stars with magnitude $<$6 and a separation of $<$3 degrees.
This indicates that these stars cannot be resolved by the lens. If we reduce the magnitude threshold for neighboring stars to $<$5, then the number of unresolved close-proximity stars drops to 26.
Furthermore, the quality of the image might be affected by the large difference in magnitude between the primary bright star and its higher magnitude neighbor(s).
The magnitudes of the halos closest to the focal point ($<$4) are on par with the magnitudes of the neighboring stars.
While we acknowledge that the proximity of two moderately low magnitude stellar neighbors is still a potential source of error in the resolving ability of the Luneburg lens, we believe that stellar proximity for high magnitude--low magnitude neighbours represent the worst case scenario for obtaining good image quality.
The Luneburg lens is likely to lose accurate data for up to 72 stars at $<$6 magnitude, which we consider to be an acceptable level of performance for all-sky imaging.

\begin{table}[!b]
\begin{center}
\begin{tabular}{lc}
\multicolumn{2}{c}{Detector specifications and Luneburg lens size} \\ \hline \hline
Pixel length & $1.75 \times 10^{-3}$ mm \\ 
Pixel resolution & $2592 \times 1944$ (5 MP) \\
CCD dimensions & 4.54 mm $\times$ 3.39 mm \\ 
Pixels/arcmin & 10 \\ 
Physical length $L$ & $1.75 \times 10^{-2}$ $\frac{\rm mm}{\rm arcmin}$  \\ 
Lens radius $r$ & 60 mm \\ \hline
\end{tabular}
\end{center}
\caption{Technical specifications of the iPhone 4 camera\cite{GSMarena,TechSpec} are used to calculate the radius of a Luneburg lens with 10 pixels per arcminute chosen as the minimum detecting quality.}
\label{tab_camera}
\end{table}

One of the considerations of manufacturing such a lens is cost.
We aim to keep lens production and implementation inexpensive.
For this reason we seek to use common consumer camera arrays as detectors to be placed on the bottom hemisphere of the lens for recording observations.
Using the technical specifications of the iPhone 4 camera\cite{GSMarena,TechSpec} as an example, we are able to approximate the physical size of the Luneburg lens. 

\begin{equation}
 \frac{L}{2 \pi r} = \frac{\theta}{360 \ \mathrm{deg}}
 \label{lens_radius}
\end{equation}

We choose a relationship between pixels and angular separation -- the example given here is 10 pixels per arcminute -- to set the limit for minimum detecting quality.
Given the pixels per arcminute and pixel edge length, we calculate the value for the physical length $L$ per arcminute on the detector.
With Equation~\ref{lens_radius} with $\theta = 1 \ {\rm arcmin} = 0.0167 \ {\rm degrees}$ we are able to calculate the maximum lens radius $r$, as seen in Table~\ref{tab_camera}.

One factor not included in this feasibility analysis is the effect of the Moon on the lens, such as image saturation during a full Moon.
A correctly modeled lunar brightness profile, which can be approximated as azimuthally symmetric, could be used to subtract away moonlight collected during a full Moon or other bright phases.
Thus, we do not consider the presence of the moon as a negative factor for observing capabilities as it can be corrected via a model lunar brightness profile.

\begin{figure}[!b]
\centering
\includegraphics[width=13cm,height=13cm,keepaspectratio=true]{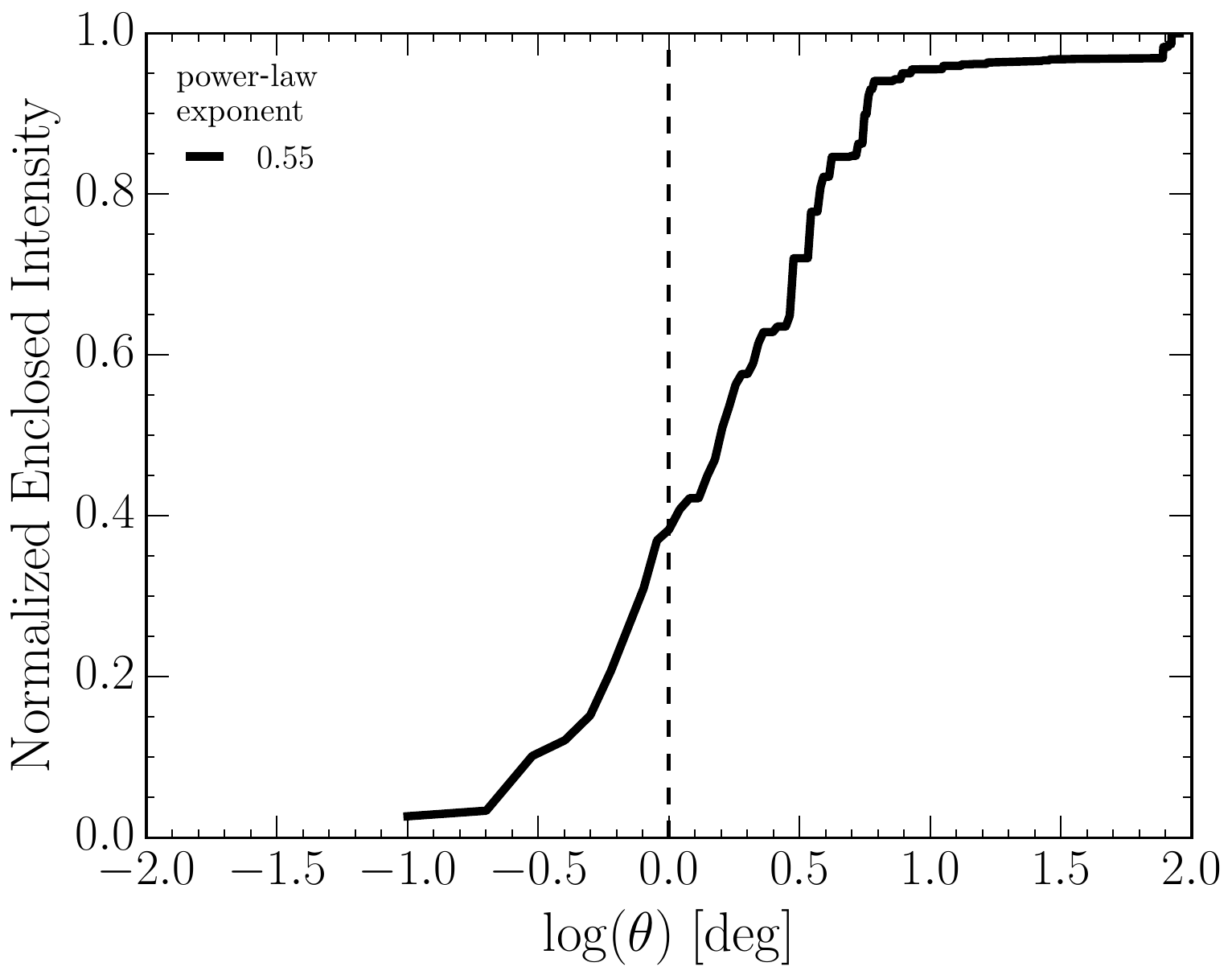}
\caption{\small{Enclosed intensity as a function of angle away from the focal point for an optimized Luneburg lens with 40 layers and a Luneburg power-law exponent of $i$ = 0.55 for 1024 rays.} }
\label{fig_encintopt}
\end{figure}

\section{CONCLUSIONS}

The stepped Luneburg lens is a compact, potentially inexpensive camera with no moving parts suitable for all-sky imaging in remote locations.
Using the \textsc{python} programming language, we develop our own ray tracing code that is able to track thousands of rays individually and record their position, direction, and amplitude upon exiting the lens.
We successfully model a stepped Luneburg lens with the ray tracing routine and optimize its parameters based on a combination for the best imaging quality.
After the analysis of different combinations of layer numbers, Luneburg power-law exponents for the refractive index, and amplitude cutoff limits, we conclude that the optimal configuration is a lens with 40 layers and a power-law exponent of $i=0.55$.
Fig.~\ref{fig_encintopt} shows the enclosed intensity as a function of angle for the optimized Luneburg lens.
Decreasing the amplitude cutoff limit does not significantly improve the efficiency or the resolution of the lens below a threshold of 1\%; increasing the amplitude cutoff to 10\% results in significantly poorer lens performance.

The full image resolution of the lens $\theta_{res} = 3.2$ degrees for the optimized Luneburg lens parameters.
Based on stellar data from the Yale Bright Star Catalog for the magnitude and proximity of bright stars, we conclude that there will be 72 stellar neighbors with magnitude $<$6 and an angular separation of $<$3 degrees that the Luneburg lens will have difficulty properly resolving.
We consider this to be an adequate level of performance for all-sky observing.
Additionally, we consider the physical limitations of the lens based on the type of camera implemented for detecting.
We take as an example the CCD chip from the iPhone 4 camera and use its specifications to determine the physical size of the Luneburg lens, which we calculate to be $r$ = 6 cm.

Most stars in the night sky  brighter than 6 magnitude could be monitored photometrically using a Luneburg lens.
The lens could use simple consumer cameras as detection devices for observations, with constraints on the size depending on the desired number of pixels per arcminute on the detector.
Factors not considered here, but potentially important, are the feasibility of fabricating a Luneburg lens with as many as 40 glass layers; potential manufacturing errors such as the decentering of lens layers or surface aberrations; and the explicit manufacturing costs of layered glass with different refractive indices.


\bibliography{report} 

\begin{thebibliography}{10}

\bibitem{Luneburg44}
Luneburg, R.~K.,  [{\em Mathematical Theory of
  Optics}{\nolinebreak\hspace{0.1em}]},  189--213, Brown University,
  Providence, Rhode Island (1944).

\bibitem{Walter60}
{Walter}, C., ``{Surface-wave luneberg lens antennas},'' {\em IEEE Transactions
  on Antennas and Propagation}~{\bf 8},  508--515 (Sept. 1960).

\bibitem{Scheel70}
{Scheel}, H.~W., ``{Experimental Investigation of a Luneberg Lens Antenna for
  Communications Satellites},'' {\em Journal of Spacecraft and Rockets}~{\bf
  7},  876 (July 1970).

\bibitem{Higgs84}
{Higgs}, J.~A. and {Worth}, R.~A., ``{The steerable Luneburg Lens as a
  communication link antenna},'' in [{\em
  ITC/USA/'84}{\nolinebreak\hspace{0.1em}]},   655--663 (1984).

\bibitem{Zernike74}
{Zernike}, F., ``{Luneburg lens for optical waveguide use},'' {\em Optics
  Communications}~{\bf 12},  379--381 (Dec. 1974).

\bibitem{Southwell77}
{Southwell}, W.~H., ``{Inhomogeneous optical waveguide lens analysis},'' {\em
  Journal of the Optical Society of America (1917-1983)}~{\bf 67},  1004--1009
  (Aug. 1977).

\bibitem{Sochacki82}
{Sochacki}, J., ``{Proposal for an alternate technology for waveguide Luneburg
  lenses},'' {\em Optics Communications}~{\bf 41},  13--16 (Mar. 1982).

\bibitem{Morgan58}
{Morgan}, S.~P., ``{General Solution of the Luneberg Lens Problem},'' {\em
  Journal of Applied Physics, Vol.~29, p.1358-1368}~{\bf 29},  1358--1368
  (Sept. 1958).

\bibitem{Hoffleit91}
{Hoffleit}, D. and {Jaschek}, C.,  [{\em {The Bright star
  catalogue}}{\nolinebreak\hspace{0.1em}]} (1991).

\bibitem{GSMarena}
GSMArena.com, ``{Apple iPhone 4 Technical Specifications}.''
  \url{https://www.gsmarena.com/apple_iphone_4-3275.php} (2018).
\newblock [Online; accessed 30-May-2018].

\bibitem{TechSpec}
TechnicalSpecifications, ``{Technical Specifications of Apple iPhone 4 Mobile
  Phone}.''
  \url{http://techspecifications.net/mobile-phones/technical-specifications-apple-iphone-4/}
  (2013).
\newblock [Online; accessed 30-May-2018].

\end{thebibliography}
\bibliographystyle{spiebib} 

\end{document}